\documentclass[]{raa}                         
\usepackage{graphicx,times}             
\usepackage{natbib}

\begin{document}

   \title{Interaction and Eruption of Two Filaments Observed by Hinode, SOHO, and STEREO}

   \volnopage{Vol.0 (200x) No.0, 000--000}      
   \setcounter{page}{1}          

   \author{Y. Li
      \inst{}
   \and M.-D. Ding$^{\star}$
      \inst{}
   }

   \institute{Department of Astronomy, Nanjing University, Nanjing 210093, China;\\
              Key Laboratory for Modern Astronomy and Astrophysics (Nanjing University),
                Ministry of Education, Nanjing 210093, China\\
             \email{dmd@nju.edu.cn}
   }

   \date{Received~~2011 month day; accepted~~2011~~month day}

\abstract{
We investigate the interaction between two filaments and the subsequent filament eruption event observed
from different view angles by Hinode, the Solar and Heliospheric Observatory (SOHO), and the Solar
Terrestrial Relations Observatory (STEREO). In the event, the two filaments rose high, interacted
with each other, and finally were ejected along two different paths. We measure the bulk-flow
velocity using spectroscopic data. We find significant outflows at the speed of a few hundreds of
km s$^{-1}$ during the filament eruption, and also some downflows at a few tens of km s$^{-1}$ at the edge
of the eruption region in the late stage of the eruption. The erupting material was composed of plasmas with a wide
temperature range of 10$^4$--10$^6$ K. These results shed light on the filament nature and the 
coronal dynamics.
 \keywords{line: profiles --- Sun: corona --- Sun: filaments --- Sun: flares --- Sun: UV radiation}}

   \authorrunning{Y. Li \& M.-D. Ding}            
   \titlerunning{Interactions and Eruptions of Two Filaments}  

   \maketitle

\section{Introduction}
\label{intro}

Solar filaments (prominences) are cold dense plasmas suspended in the corona. Their fine threads are
seen in emission as bright prominences at the limb and in absorption as dark filaments against the disk.
A filament is formed and maintained above the magnetic polarity inversion line, in a magnetic
structure called a filament channel, in which the filament can be supported by the magnetic field. 
For filament formation and maintenance, \citet{mart98} gave a comprehensive review,
including the filament structure, chirality, magnetic topology, and mass flows. Filaments would erupt
at the onset of MHD instabilities or loss of equilibrium. It is now widely accepted that filament eruption 
is often associated with other solar activities, such as solar flares and coronal mass ejections (CMEs).
\citet{martens89} considered filaments as the coronal part of an electric
current that loses MHD equilibrium at the flare onset and starts to erupt outwards. This leads directly
to the observed CME. \citet{kaas85} showed that as the filament moves upwards, a neutral line is
formed beneath it, which becomes the site of magnetic reconnection and particle acceleration 
during the flare. The accelerated particles precipitate into the chromosphere along the field lines, 
and produce hard X-ray, UV, and optical (such as H$\alpha$) emission forming flare ribbons. The energy deposition
in the lower atmosphere further causes chromospheric evaporation, which leads to the observed coronal extreme-ultraviolet
(EUV) and soft X-ray emissions \citep{canf82}.

It has been proposed that two filaments (or filament segments) of the same chirality, when approaching each other, 
can merge to form long filaments or interact with each other \citep{daza48, martens01}. This has been confirmed by
observations \citep{schm04} and numerical simulations \citep{devo05, aula06}. However, in these observations
and numerical experiments, the authors found no initiation of an eruption. Theoretically, 
\citet{martens01} presented a ``head-to-tail'' linkage model for the eruption of filaments. Furthermore,
\citet{ural02} suggested that the dual-filament interaction could cause solar eruptions. It is 
fortunate that there are indeed a few observations indicating that
the linkage of two filaments can initiate an eruption. \citet{sujt07} presented new observations
of the interaction of two nearby but distinct H$\alpha$ filaments and their successive eruptions. In
the event, the interaction was initiated mainly by an active filament of them. They considered that the
second filament eruption may be the result of a loss of stability owing to the sudden mass injection
from the first filament eruption. \citet{liuy10}
presented another interesting case: two filaments erupted simultaneously and there was no transfer of
material between them during the initial stage; the two filaments merged together along the ejection path,
indicating the coalescence between the two interacting flux ropes. Moreover, \citet{bone09}
reported observations of the interaction and merging of two filaments, one active and one quiescent,
and their subsequent eruption. Even so, observations of interacting and erupting filaments are still rare at present.

With the multi-wavelength observations from Hinode, the Solar and Heliospheric Observatory (SOHO),
and the Solar Terrestrial Relations Observatory (STEREO), we investigate the interaction and eruption
of two filaments. We present observations of the filament interaction and mass ejection, and also measure
the bulk-flow velocities of the eruption using spectroscopic data. It is found that the two filaments
rise to approach each other, interact, and finally are ejected along two different paths. We describe the
observing instruments and data in Section \ref{obs}. The evolution of the filament interaction and
eruption is presented in Section \ref{picture}. Section \ref{erup_EIS} gives the bulk-flow velocities
of the eruption. Finally, a discussion is given in Section \ref{discussion}.

\section{Overview of the Observations and Instruments}
\label{obs}

\begin{figure}
\begin{center}
\includegraphics[width=0.9\textwidth]{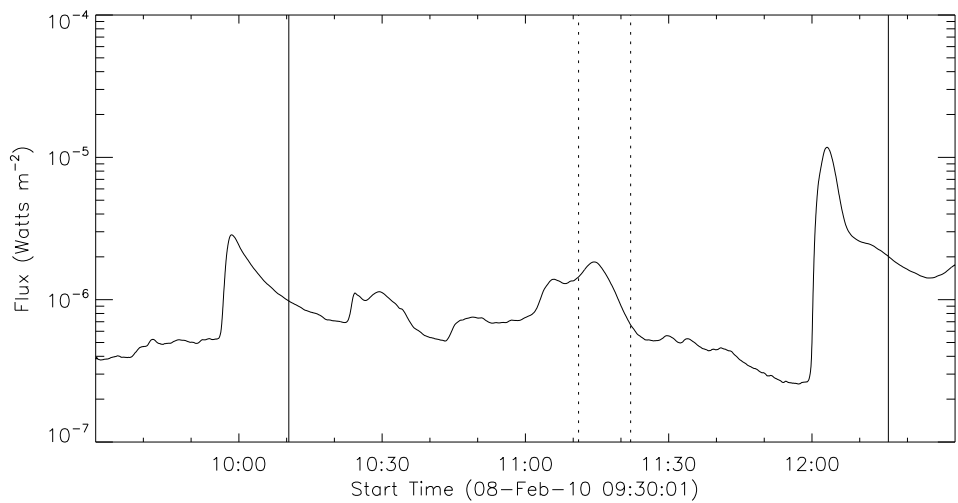}
\end{center}
\caption{GOES 1--8 \AA~soft X-ray light curve for a period covering the filament interaction and eruption.
The vertical solid lines show the time range of the EIS scan over the active region,
while the dotted lines show the time period of filament eruption.}
\label{goes_flux}
\end{figure}

The event of filament eruption was located in the solar active region NOAA 11045. The active region
produced several C-class and an M-class flares within a few hours on 2010 February 8. Among them, 
a GOES C1.8 flare, which started at 11:00 UT and peaked at 11:14 UT, was in close association with 
the filament eruption. Figure~\ref{goes_flux} shows the GOES 1--8 \AA~soft X-ray light curve around the flare time. In addition,
a white-light CME was first detected with SOHO/LASCO at 06:30 UT, having a linear speed of 153 km s$^{-1}$ and lasting about 12 hours.

\subsection{Optical Images from Hinode/SOT}

The filament eruption was observed in H$\alpha$ images taken with Hinode/SOT (Solar Optical Telescope;
\citealt{tsun08}) with a spatial resolution of 0.16$''$ per pixel. SOT observed the event in
H$\alpha$ every half a minute before 10:22 UT and after 12:03 UT, but only obtained one image in between
(at 11:41 UT). During this period, SOT was observing the same active region using the Ca II H line with
a cadence of ten minutes; thus we can study the evolution of the two flare ribbons. SOT filtergraph
magnetograms are not available for this study. Instead, we use the data from SOHO/MDI (Michelson Doppler
Imager; \citealt{sche95}) to study the magnetic field in this active region.

\subsection{EUV Images from SOHO/EIT and STEREO/SECCHI/EUVI}

The interaction and eruption of the filaments were observed in EUV by SOHO/EIT (Extreme-ultraviolet
Imaging Telescope; \citealt{dela95}) and by the SECCHI/EUVI (Sun Earth Connection Coronal and Heliospheric
Investigation/Extreme UltraViolet Imager; \citealt{wuls04}) instrument onboard the STEREO spacecraft
(A and B). The EIT images were obtained in 195 \AA~($\sim$1.5$\times$10$^6$~K) with a cadence of 12 minutes,
while the SECCHI/EUVI images were observed in 304 \AA~($\sim$6.0$\times$10$^4$~K) and 195 \AA~with
cadences of 10 and 5 minutes, respectively. The active region was located near the center of the solar
disk in the EIT field of view (FOV), while it appeared on the east and west limbs in the STEREO-A and B FOVs,
respectively. On that day, these two spacecraft were separated by an angle of $\sim$136$^{\circ}$. We
use the EUV data between 10:40 UT and 11:30 UT. This period is from about 20 minutes prior to the flare
onset to about 30 minutes after it, thus covering the entire eruption process.

\subsection{Spectroscopic Data from Hinode/EIS}

We measure the Doppler velocities of the mass ejection from spectroscopic data taken with Hinode/EIS
(Extreme-ultraviolet Imaging Spectrometer; \citealt{culh07}). EIS covers two wavelength bands, 170--211
\AA~and 246--292 \AA, referred to as the short and long wavelength bands, respectively. These bands
include some strong transition region and coronal lines over a wide temperature range. We mainly analyze
the Fe {\sc xii} 195.12 \AA~line here. In observations, the 2$''$ slit of EIS was used to scan
the active region of 324$''\times$376$''$ with an exposure time of 45 s. The scanning
lasted for 2 hours and 6 minutes, as indicated in Figure \ref{goes_flux}.
Our study focuses on the region hosting the eruption, which has a size of about
30$''\times$90$''$. EIS slit crossed this region from 11:11 UT to 11:22 UT
(between the two vertical dotted lines in Figure \ref{goes_flux}). Note that, for all the images from EIS,
we use the coordinates in arcseconds relative to a reference point inside the EIS raster FOV.

We reduce the EIS data using the standard processing package. This corrects the detector bias and dark
current, as well as hot pixels and cosmic ray hits, resulting in absolute intensities in
erg~cm$^{-2}$~s$^{-1}$~sr$^{-1}$~\AA$^{-1}$. We also make a correction for a slight tilt of the slit
on the CCDs. In addition, we correct for the variation in spectral line positions over the Hinode
orbit caused by temperature variations in the spectrometer. Such an orbital variation is obtained by
averaging the centroid positions over the length of the slit for a quiet region. When measuring the
relative Doppler velocity, we use the average line center over a quiet region as the reference Fe {\sc xii} 
wavelength.

To interpret the multi-wavelength observations, it is important to co-align the different images.
There is an instrumental offset between the images taken in the two EIS CCDs \citep{youn07a}.
To correct this offset, we shift the long wavelength images by 2$''$ in the solar X-direction
and 17$''$ in the solar Y-direction. To co-align the images from SOT and MDI, we use the
sunspots observed by both instruments. The co-alignment between the EIT and EIS data
is made using the 195 \AA~images obtained by both instruments. We estimate the uncertainty 
of the co-alignment to be within 5$''$--10$''$.

\section{Filament Interaction and Eruption}
\label{picture}

\subsection{The Filaments and Associated Flare}
\label{ejec_SOT}

\begin{figure}
\centerline{\includegraphics[width=1.0\textwidth]{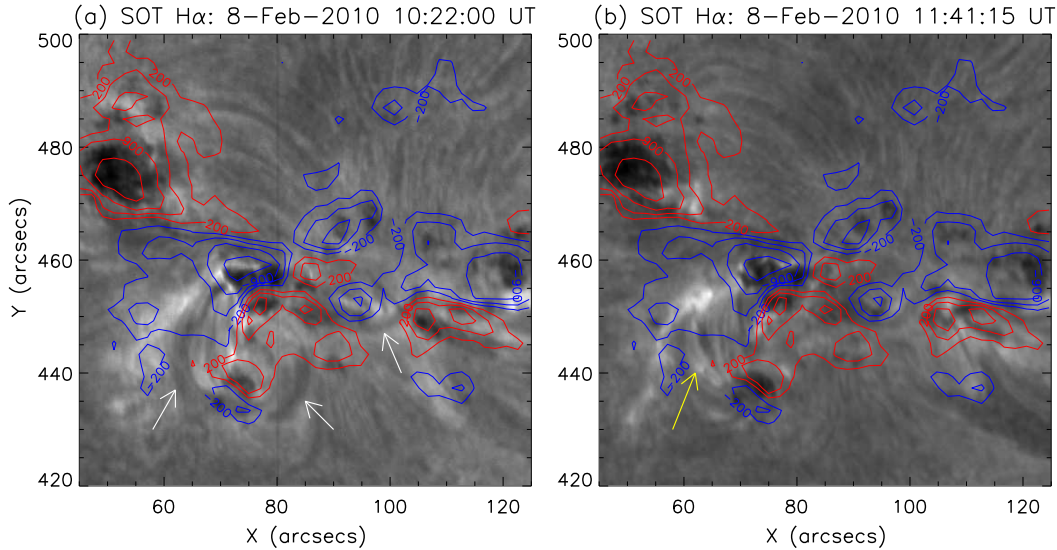} }
\caption{Hinode/SOT~H$\alpha$ images before and after the flare onset at 11:00 UT,
superimposed with the MDI magnetogram taken at 11:15 UT on February 8. Red and blue
contours refer to the positive and negative polarities, respectively, with levels of
200, 500, 900, and 1500 G. The white arrows in panel (a) point to the two close filaments
existing prior to the flare onset. The yellow arrow in panel (b) marks the post-flare loops.}
\label{Ha_MDI}
\end{figure}

Figure \ref{Ha_MDI} shows the SOT H$\alpha$ images of the flaring region overlaid with the contours
of the MDI magnetogram. The two H$\alpha$ images were observed $\sim$40 minutes prior to and after
the flare onset at 11:00 UT, respectively. The superimposed MDI magnetogram was taken at 11:15 UT,
with contour levels of 200, 500, 900, and 1500 G. There are some strong positive (red contours)
and negative (blue contours) magnetic patches. The white arrows (Figure \ref{Ha_MDI}a) mark the
two filaments prior to the flare onset. 
Considering the filament morphology and the magnetic configuration, it is reasonable to speculate 
that the dark segments indicated by the two left arrows are magnetically connected with each other, or 
that they are the two segments of one large filament. In this context, we only mention one filament, 
which includes the two segments. 
After the flare (Figure \ref{Ha_MDI}b), the two filaments disappeared, and some post-flare loops 
(shown by the yellow arrow) appeared instead. From the two H$\alpha$ images, we find that the filaments 
erupted between 10:22 UT and 11:41 UT. 
The interaction between the left and right filaments and their eruption 
are more clearly presented in the EUV images described in Section \ref{erup_EUV}.

\begin{figure}
\centerline{\includegraphics[width=1.0\textwidth]{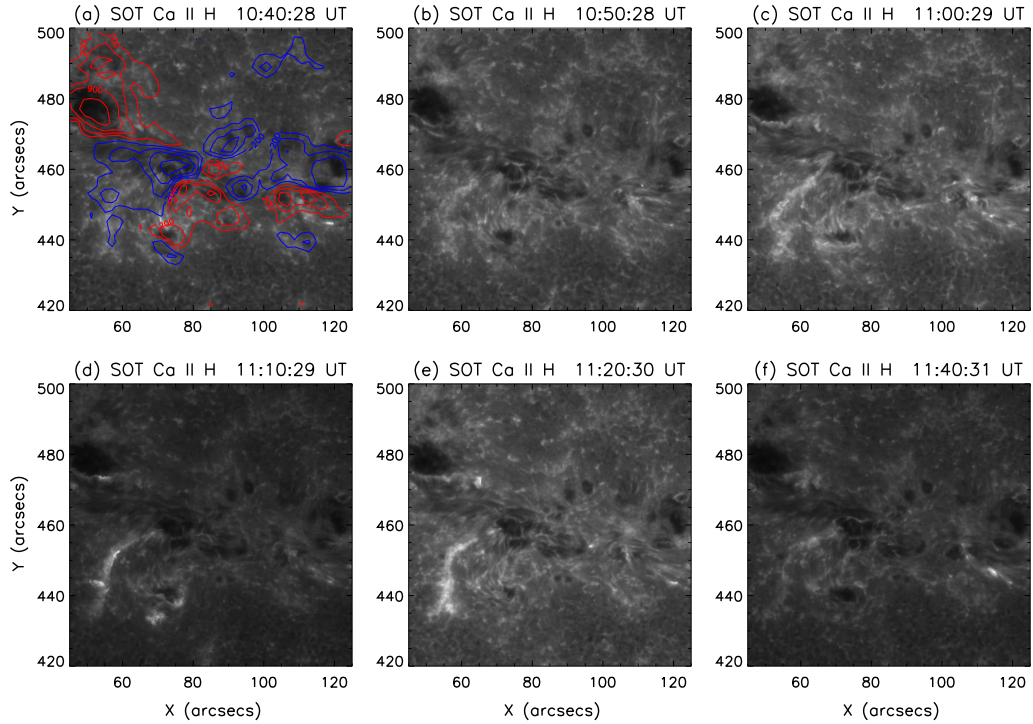} }
\caption{SOT~Ca II H images during the flare. The image in panel (a) is overlaid with
the MDI magnetogram taken at 11:15 UT. The MDI contours and the FOV are the same as
that in Figure \ref{Ha_MDI}.}
\label{CaIIH}
\end{figure}

Figure \ref{CaIIH} shows the evolution of the two flare ribbons. The SOT Ca II H images have the
same FOV as the H$\alpha$ images in Figure \ref{Ha_MDI}. The left filament was located 
between the two flare ribbons, and the right one was on the right side of the ribbons. 
The two flare ribbons began to brighten
at $\sim$11:00 UT (Figure \ref{CaIIH}c). They became the brightest near the peak time of the flare soft X-ray emission
(Figure \ref{CaIIH}d). Tens of minutes later, the ribbons gradually disappeared (Figures \ref{CaIIH}e
and \ref{CaIIH}f).

\subsection{The Filament Interaction and Eruption}
\label{erup_EUV}

\begin{figure}
\centerline{\includegraphics[width=0.7\textwidth]{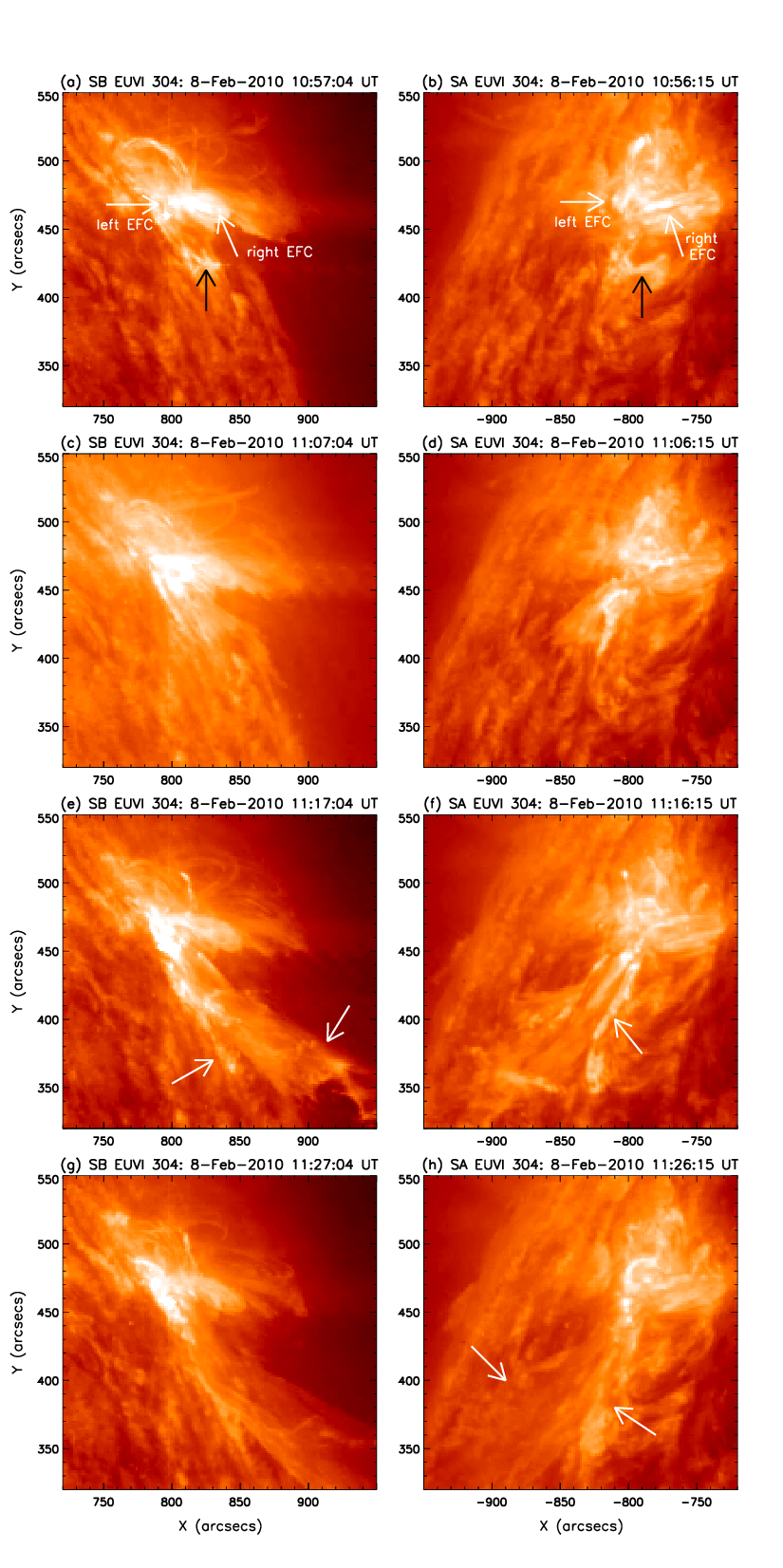} }
\caption{SECCHI/EUVI 304~\AA~images of the filament eruption region (evolving from top to bottom). 
``SA'' refers to STEREO-A and ``SB'' to STEREO-B. The black arrows in (a) and (b) mark the bright
expanding loops. The white arrows point to the filaments or EFCs.}
\label{EUVI304}
\end{figure}

\begin{figure}
\centerline{\includegraphics[width=1.0\textwidth]{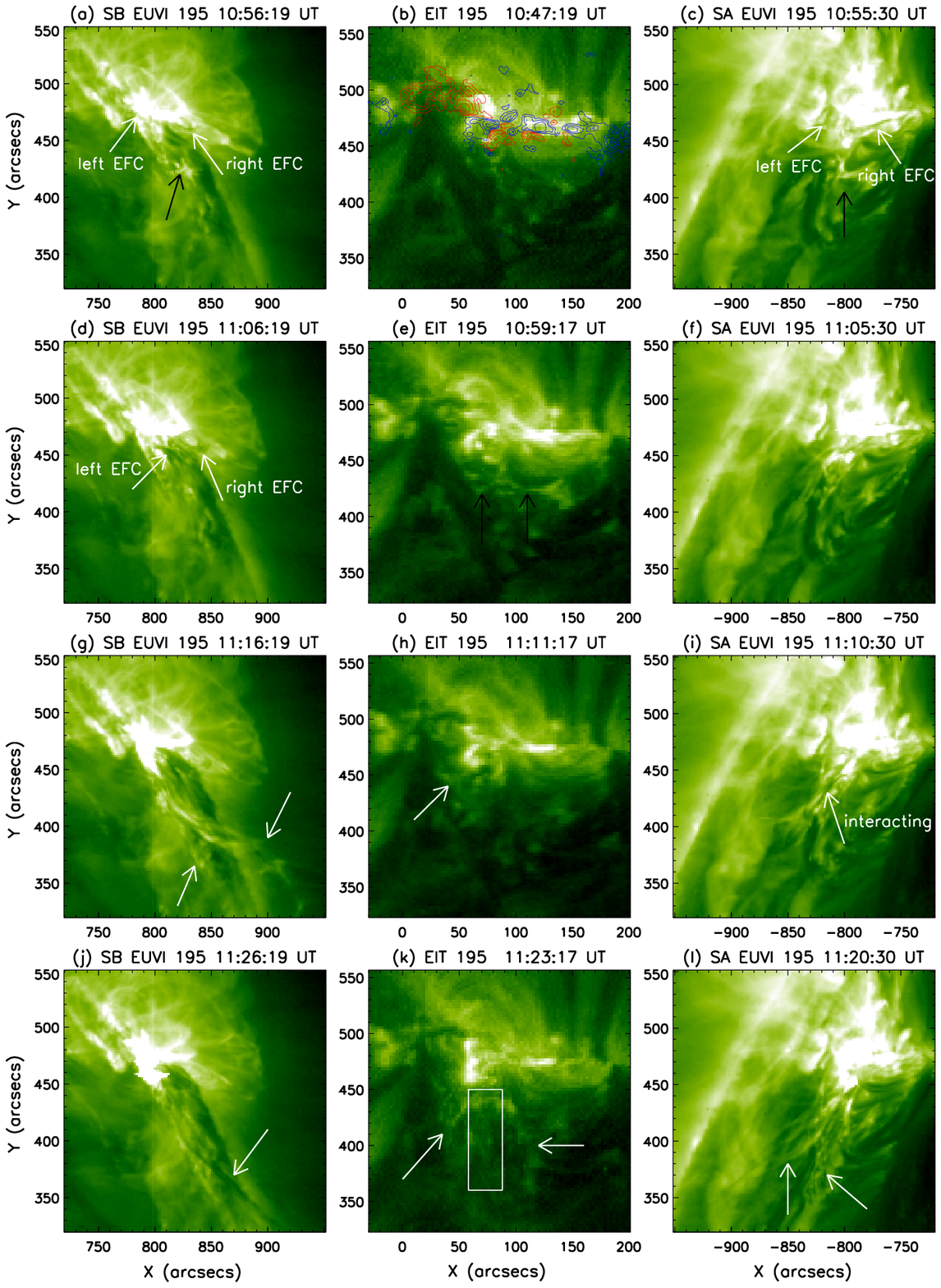} }
\caption{EIT 195 \AA~images and EUVI 195 \AA~images of the filament eruption region (evolving from top to bottom). 
``SA'' and ``SB'' refer to STEREO-A and STEREO-B, respectively. In (b), the EIT image
is overlaid with the MDI magnetogram taken at 11:15 UT. The black arrows in (a), (c),
and (e) mark the bright expanding loops. The white arrows point to the filaments or EFCs. 
In (k), the white box shows the filament eruption region that we focus on in the study.}
\label{195all}
\end{figure}

The filament interaction and eruption are most evident in EUV images from EIT
and SECCHI/EUVI. In the FOVs of these instruments, the eruptions appeared near the center
of the solar disk and on the east and west limbs, respectively, as mentioned above. Therefore,
we can view this event in three different perspectives. Figures \ref{EUVI304} and \ref{195all}
give the pictures in the filters of 304 \AA~and 195 \AA. Prior to the onset of the flare,
we can see the loop expansion due to the filament rising. There appeared some bright loop
structures (marked by the black arrows in Figures \ref{EUVI304}a, \ref{EUVI304}b, \ref{195all}a,
\ref{195all}c, and \ref{195all}e) at $\sim$10:56 UT below the core region as compared with
the structure at an earlier time (Figure \ref{195all}b). In particular, there were some loops
being contacted (Figure \ref{195all}e). This may lead to or accelerate the filament interaction
and eruption. 
The two EUV filament channels (EFCs) can be seen in Figures \ref{EUVI304}a, \ref{EUVI304}b,
\ref{195all}a, \ref{195all}c, and \ref{195all}d, as marked by the white arrows. 
At an early time ($\sim$10:56 UT), both of the filaments were relatively stable. 
They were rising higher and about to erupt at $\sim$11:06 UT; 
meanwhile, the two flare ribbons were brightened (see Figure \ref{CaIIH}). 
A few minutes later, the two filaments collided and began to erupt,
as shown in Figure \ref{195all}i. Due to the low temporal and spatial resolutions, we 
cannot distinguish which segment (or both segments) of the left filament met with
the right filament. However, we can judge that the right EFC made a 
collision with the left EFC.  This was also confirmed by the observation
that the ejected material from the filament suddenly changed
the direction of ejection: the cool material was expelled to
the left side as shown in Figure \ref{195all}h when the two EFCs came into contact 
with each other (shown more clearly in the video 1, available in the electronic edition
of the paper). It seems that the two filaments erupted nearly simultaneously
after the peak of the flare soft X-ray emission at 11:14 UT (Figures \ref{EUVI304}e--h, \ref{195all}g, and
\ref{195all}l). Moreover, a careful inspection reveals two ejection
paths, marked by the two arrows in Figures \ref{EUVI304}e, \ref{EUVI304}h, \ref{195all}g,
and \ref{195all}l. Because of the projection effect, the two ejection paths are not clearly visible
in EIT images (Figure \ref{195all}k). The eruptions along different directions
can be seen more clearly in the movies (videos 2 and 3). At the later stage of the eruption, some of
the filament material cooled down (Figure \ref{195all}j) and fell back to the solar surface,
while the remainder was ejected into the space, as judged from the ejection velocities measured
in Section \ref{erup_EIS}.

From different view angles (Figure \ref{195all}), we see the different manifestations of
filament eruption. Near the solar disk center, the filament eruption appears rather diffuse; while
at the limbs the observed ejection is rather collimated. Furthermore, we observe the different
shapes of the erupting filaments from different passbands. In the 304 \AA~image, the filament
ejection is more extended; while in the 195 \AA~image, the ejected material is more confined. 
The ejection being visible in emission in a few passbands suggests that the filament material
is composed of plasmas with a wide temperature range from 6.0$\times$10$^4$~K to 1.5$\times$10$^6$~K
(see also Figure \ref{EISinten}). It is possible that part of the filament material is heated to the 
coronal temperature during the interaction and eruption. An alternative view is that the ejection
contains hot plasmas from the reconnection outflow. In a word, cool material ($\sim$10$^{4}$~K) and 
hot material ($\sim$10$^{6}$~K) co-exist in the eruption.

\section{Bulk-flow Velocities of the Eruptions}
\label{erup_EIS}

\begin{figure}
\centerline{\includegraphics[width=1.0\textwidth]{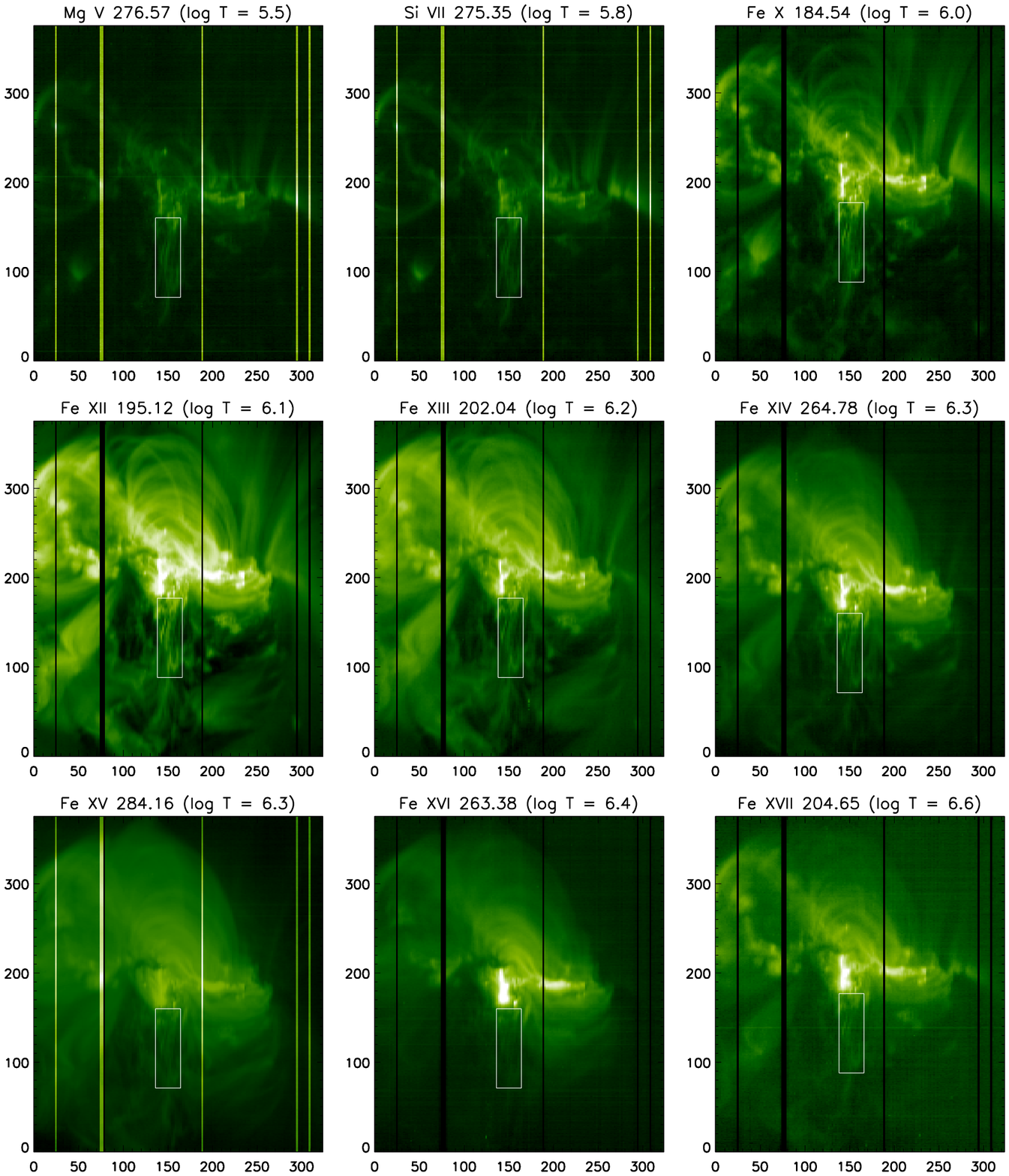} }
\caption{Intensity maps of the EIS spectral lines. The white boxes mark the main eruption
region with the same FOV as the white box in Figure \ref{195all}. The filament material is visible
in most of these images. Note that the coordinates are relative to a reference point in
arcseconds inside the EIS raster FOV (the same for Figures \ref{EIS195} and \ref{EIS195part}),
and that there is an instrumental offset between the two EIS CCDs recording the short and
long wavelength images, respectively. The dark/bright vertical lines refer to missing/abnormal
data points.}
\label{EISinten}
\end{figure}

The EIS spectroscopic data were obtained by scanning the active region from solar west to east
between 10:10 UT and 12:16 UT.
Figure \ref{EISinten} shows the intensity maps in EIS spectral lines. The white boxes mark the
main eruption region, which was scanned from 11:11 UT to 11:22 UT. During that time, the filaments
were erupting. We can see the bright emission from the flare kernel and from the
filament eruption in most spectral lines. The filament material is visible in the low temperature line
of Mg {\sc v}~($\sim$3.0$\times$10$^5$~K), as well as in the high temperature line of
Fe {\sc xv}~($\sim$2.0$\times$10$^6$~K).

\begin{figure}
\centerline{\includegraphics[width=1.0\textwidth]{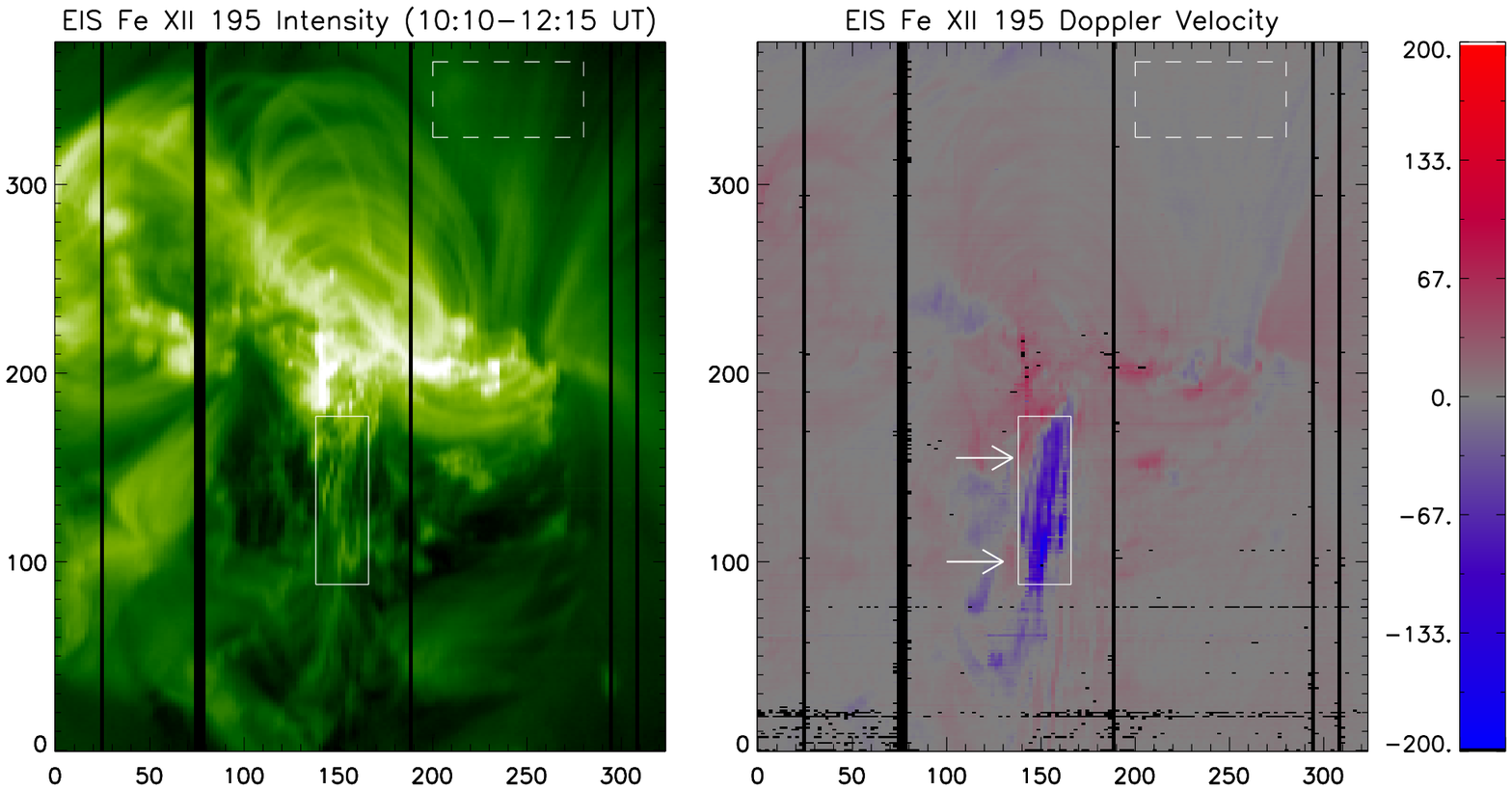} }
\caption{EIS 195~\AA~intensity and Doppler velocity maps. The white solid box marks the
filament eruption region. The white dashed box represents the quiet region that we use to
calculate the reference wavelength. All the values in the velocity panel are obtained
from the single Gaussian fitting. The arrows point to the downflow areas at the edge
of the main eruption region.}
\label{EIS195}
\end{figure}

\begin{figure}
\centerline{\includegraphics[width=1.0\textwidth]{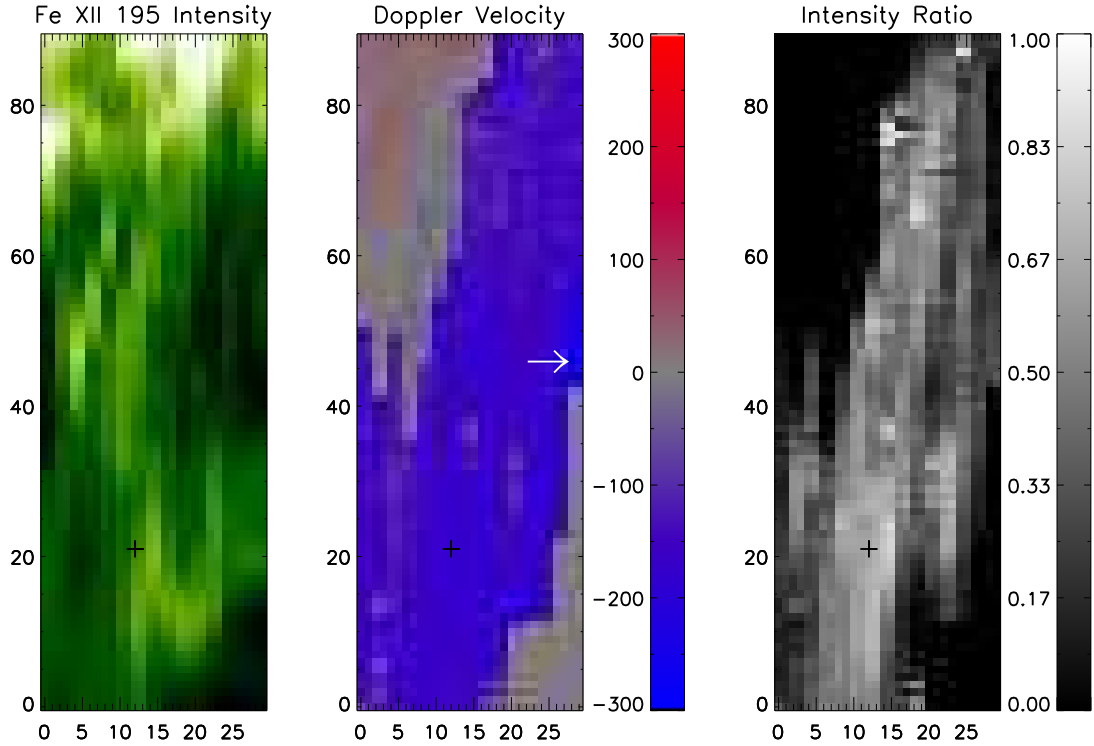} }
\caption{EIS 195~\AA~intensity and Doppler velocity maps of the filament eruption region
marked by the white solid box in Figure \ref{EIS195}. The velocity field is obtained by
combining the results from either the double Gaussian fitting or the single Gaussian fitting.
When the double Gaussian fitting is applied, the velocities refer to those of the blueshifted
components. The right panel shows the intensity ratio between the blueshifted component and
the whole profile. The ratio equals zero when the profile can be well fitted by a single
Gaussian function. The white arrow points to the region showing large Doppler velocities over
250 km s$^{-1}$ at the beginning of the eruption. The line profile and its fitting result
at the plus sign is shown in Figure \ref{profile}.}
\label{EIS195part}
\end{figure}

\begin{figure}
\centerline{\includegraphics[width=0.8\textwidth]{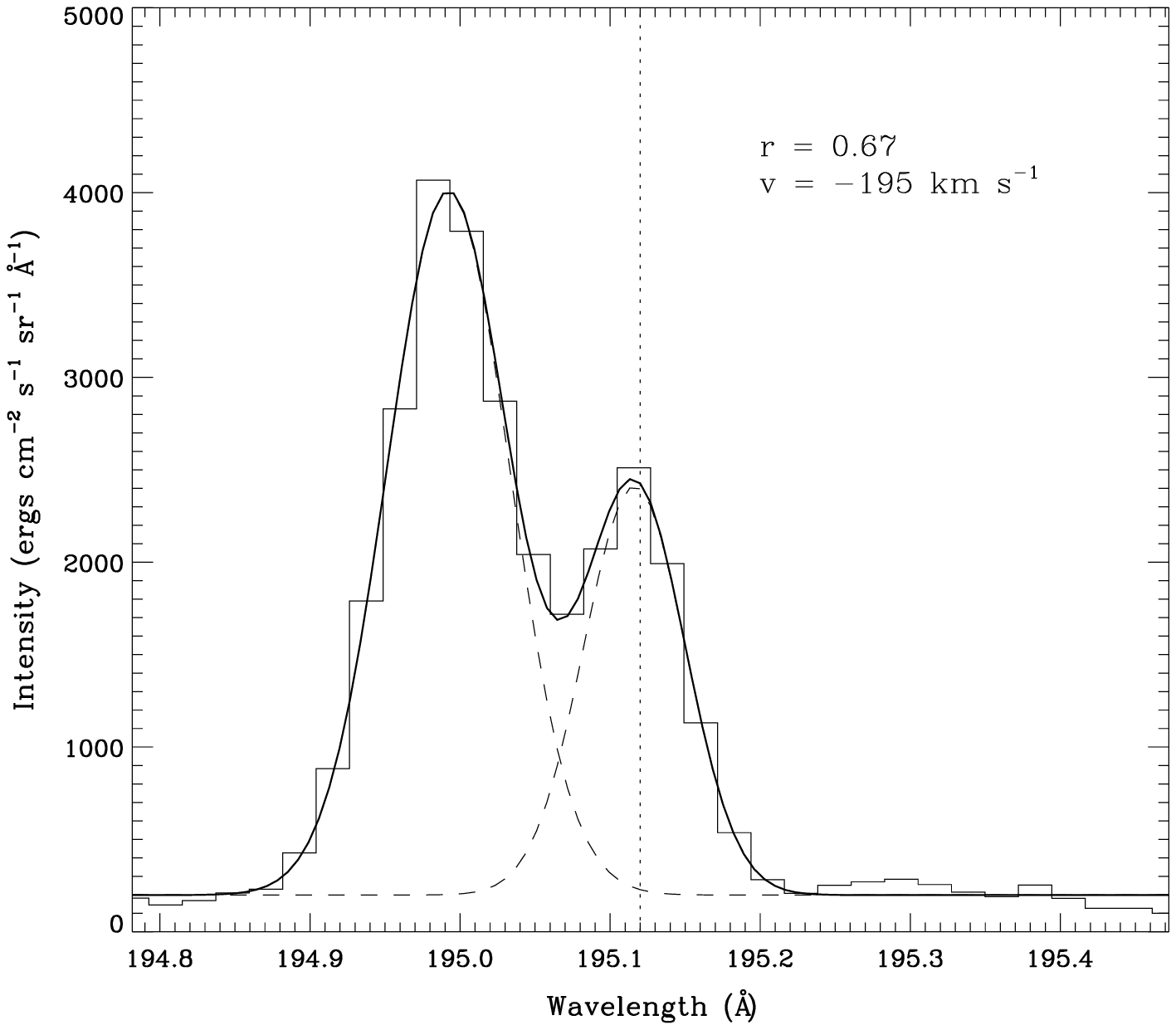} }
\caption{A typical line profile and its double Gaussian fitting at the plus sign in Figure
\ref{EIS195part}. The histogram is the observed profile and the solid line is the fitting
result. The vertical dotted line refers to the reference wavelength. This line is well
fitted by double Gaussian components that are plotted with dashed curves. The velocity
for the blueshifted component is 195 km s$^{-1}$, and its intensity ratio to the whole
profile is 0.67.}
\label{profile}
\end{figure}

Figure \ref{EIS195} shows the Fe {\sc xii} 195.12 \AA~intensity and Doppler velocity maps.
In this figure, the Doppler velocity is derived from a single Gaussian fitting to the observed
spectral line. We detect significant outflows (negative velocities) with a speed of hundreds 
of km s$^{-1}$ in the eruption region. In addition, the analysis also reveals some weak redshift 
(positive velocities) of about tens of km s$^{-1}$ near the edge of the region (indicated by the arrows). 
The redshifts may originate from the cooling plasmas, which were falling back to the solar surface
at the late stage of the eruption when the EIS slit passed over the edge of the
region. When measuring the Doppler velocity, we notice that, in the quiet and redshifted regions,
the line profiles can be well fitted by a single Gaussian function; while in the region with
significant outflows, the line profiles show obvious double Gaussian components, with a blueshifted
component superimposed on a relatively static one. Therefore, we consider to use the double Gaussian
function to fit the line profiles in the main eruption region. Figure \ref{EIS195part} shows
the results of the Fe {\sc xii} line. In the velocity panel, most of the numbers are
derived from a double Gaussian fitting. We also calculate the ratio \textit{r},
of the integrated intensity of the blueshifted component to that of the entire profile.
It is plotted in the ratio panel (\textit{r} becomes zero when the line profile is well
fitted by a single Gaussian function). We find that at many locations the blueshifted
component is stronger than the static component, i.e., the intensity ratio as defined above
is over 0.5. Figure \ref{profile} shows an example of double Gaussian fitting for the pixel
marked by the plus sign in Figure \ref{EIS195part}. The intensity of the blueshifted component
reaches even 67\% of the total intensity, and the blueshifted velocity is 195 km s$^{-1}$.
Double Gaussian shapes are also present in other line profiles (such as Fe {\sc xiii} and Fe {\sc xv}).
In the region marked by the arrow in the velocity panel of Figure \ref{EIS195part}, the velocity
of the blueshifted component can be as high as 300 km s$^{-1}$, though its intensity is less
than half of the total. If corrected for the projection effect, the true velocity may be
greater than the escape velocity on solar surface ($\sim$618 km s$^{-1}$). 
It means that there is some material
going into the space as mentioned in Section \ref{erup_EUV}. In fact, the time when the EIS
slit scanned over this region, $\sim$11:11 UT, corresponds to the onset of the filament eruption.

The spectral line profile in the eruption region cannot be fitted by a single Gaussian function.
Although the low spatial resolution of the scanning observation may cause such a deviation from Gaussian, 
we think it more likely that there are multi-velocity components along the line of sight. 
This is supported by the observation that the two filaments are ejected along different trajectories.

\section{Discussion}
\label{discussion}

We study the interaction and eruption of two filaments observed from three different view angles.
The event was accompanied by a C1.8 two-ribbon flare. At first, we observe the expansion of coronal loops.
Then, the two filaments rose higher, interacted with each other, and were finally ejected
along two different paths. During the interaction and eruption, the flare ribbons were brightened
and then gradually decay. The filament eruption was observed in emission in both the 304 \AA~image and the 195 \AA~band,
suggesting that the erupted filament material contained plasmas with a wide temperature range
of 10$^{4}$--10$^{6}$ K. We measure the bulk-flow velocities using the EIS spectroscopic data. Significant 
outflows were detected with a speed of several hundred km s$^{-1}$ during the eruption. Downflows of tens 
of km s$^{-1}$ were also observed at the edge of the eruption region during the late stage of the eruption. 
Most of the blueshifted line profiles are double-peaked and can be fitted to two gaussians. After correcting 
for the projection effect, we find that the upward velocity may exceed the solar
escape velocity, suggesting that some filament material may be ejected into the space.

This event occurred in an active region that produced a number of large flares. From
the movie of the MDI 96-min magnetogram (video 4, available in the electronic edition), we can
see that the strong positive and negative magnetic patches were approaching, disintegrating,
canceling, and disappearing, especially in the flaring region at $\sim$11:00 UT. 
From the available data we find that there is a close relationship between the flare and the
filament eruption. Before the flare started, the coronal loops appeared to expand. Then the filaments
rose higher. Several minutes later, the flare ribbons were brightened. 
It is therefore likely that magnetic reconnection occurred between the two filaments and triggered the filament eruption.
As the filaments rose, they began to be linked together, interact with each other, and were finally erupted. After the
peak of the flare, the filaments were ejected along two different paths. Some of the material escaped
from the Sun, and the remainder cooled down and fell back to the solar surface. The interaction of the
two filaments may lead to or accelerate the flare energy release.

It has been discussed that pairs of filaments (or filament segments) can merge and interact with each other
\citep{daza48, mart98, martens01, schm04, devo05, aula06}. However, these studies are mainly focused on
the filament formation. Observations of the evolution and eruption of filament pairs are
valuable and can provide a better understanding of the filament nature and coronal dynamics.
To our knowledge, \citet{sujt07} reported the first observational study of the interaction
between two distinct H$\alpha$ filaments and the successive eruptions. In their event, the two
filaments erupted with an interval of 40 minutes. It is observed that some material was transferred
from the filament that erupted first to the other filament, and triggered its eruption. 
This event was associated with a CME. Quite recently, \citet{liuy10} presented another
interesting event: two filaments erupted simultaneously and there was no mass transfer
between them during the initial stage. The two filaments merged together along the ejection
path and no CME was observed. Furthermore, \citet{bone09} reported observations of
the interaction and full merging of two filaments, one active and the other quiescent, and their
subsequent eruption. In these observational studies, the authors mainly used
the ground-based H$\alpha$ data with high cadence that observes filament plasmas at a low temperature.
In our study, however, we use multi-wavelength images in both EUV and H$\alpha$ to show
the interaction and eruption of the two filaments from three different  view angles. We can
compare this event with the previously reported events. The two filaments in this event seem to erupt 
simultaneously without evident mass transfer between them. This is similar to the case of
\citet{liuy10}, but different from the one studied by \citet{sujt07}.
Moreover, our event was associated with two active filaments, while the event presented by 
\citet{bone09} involved one active filament and one quiescent filament. In particular, we find
that the two filaments were ejected along two different paths. These different paths of ejections 
can be explained by the 3D magnetic reconnection as shown in numerical simulations of 
\citet{jiang11a, jiang11b}. The magnetic reconnection occurred when the two filaments were rising and interacting with each other. 
Moreover, in observations, the EFCs did not seem to merge; therefore, 
they were linked only partially before the filament material was ejected. This may also have some relation 
to the different ejection paths. The details need to be verified by numerical simulations. 

The chirality (handedness) of filaments plays an important role in the
merging and linkage of filament pairs. Our results support the scenario that 
only filaments of the same chirality can link up \citep{martens01, schm04, devo05, aula06,sujt07}. 
In our event, the two filaments are both dextral, and they are linked up and interact with each other. This 
is consistent with the previous research.

Spectroscopic data are important for the study of filament heating and coronal dynamics \citep{ying09}. In this event, 
we only analyze the EIS Fe {\sc xii} 195.12 \AA~line to obtain the bulk-flow velocity of the filament eruption.
In fact, the EIS line profiles are complicated in most regions, especially in eruption regions.
They are often double-peaked, consisting of static and shifted components \citep{ying11}, which
can be well fitted by double Gaussians. The double-peaked profile may indicate presence of multiple line-of-sight
velocities or may result from the low spatial resolution of the observation.
In either case, the dynamics of the filaments and corona are very complicated during the eruption. 
In addition, we can use the spectroscopic data to measure the temperature and density of filaments and 
to analyze the plasma heating and cooling during the eruption. To this end, we need spectroscopic 
data and multi-wavelength images with higher temporal and spatial resolutions.

\begin{acknowledgements}
The authors would like to thank P.~F. Chen, J. Qiu, and Y. Guo for their helpful discussions 
and the referee for valuable comments on the paper. This work
was supported by NSFC under grants 10828306 and 10933003 and by NKBRSF under grant 2011CB811402.
Hinode is a Japanese mission developed and launched by ISAS/JAXA, collaborating with NAOJ as a
domestic partner, and NASA (USA) and STFC (UK) as international partners. Scientific operation
of the Hinode mission is conducted by the Hinode science team organized at ISAS/JAXA. Support for
the post-launch operation is provided by JAXA and NAOJ (Japan), STFC (U.K.), NASA, ESA, and NSC 
(Norway).
\end{acknowledgements}


\end{document}